\documentclass[10pt, conference]{IEEEtran}
\IEEEoverridecommandlockouts
\usepackage{graphicx}
\usepackage{svg}
\usepackage{comment}
\usepackage{bbding}
\usepackage{multirow}
\usepackage{longtable}
\usepackage{array}
\usepackage{url} 
\usepackage{cite}
\usepackage{float}
\usepackage{amsmath}
\usepackage{amssymb}
\usepackage{indentfirst}
\usepackage{verbatim}
\usepackage{makecell}
\usepackage{soul}
\usepackage{tabularx}
\usepackage{booktabs}
\usepackage{subcaption}
\usepackage{lettrine}
\usepackage{soul}
\usepackage{algorithm}
\usepackage{algpseudocode}
\usepackage{colortbl} 

\usepackage[shortcuts,acronym, nonumberlist]{glossaries}

\newcommand{\Rev}[1]{{\color{blue}{#1}}}
\def\BibTeX{{\rm B\kern-.05em{\sc i\kern-.025em b}\kern-.08em
    T\kern-.1667em\lower.7ex\hbox{E}\kern-.125emX}}

\makeglossaries 

\newacronym{uwb}{UWB}{{Ultra-Wideband}}
\newacronym{cir}{CIR}{{Channel Impulse Response}}
\newacronym{los}{LOS}{{Line-of-Sight}}
\newacronym{nlos}{NLOS}{{Non-Line-of-Sight}}
\newacronym{ssm}{SSM}{{state space model}}
\newacronym{lut}{LUT}{{lookup table}}
\newacronym{mae}{MAE}{{mean absolute error}}
\newacronym{mocap}{MoCap}{{motion capture}}
\newacronym{mse}{MSE}{Mean Squared Error}

\begin{document}

\title{Tracking the Turn: Mamba-Powered Human Orientation Detection using UWB\\
\thanks{This research was partly funded by the imec ICON project NG-UWB (Agentschap Innoveren en Ondernemen project nr. HBC.2024.0882), the Fund for Scientific Research Flanders (FWO-Vlaanderen) under SB-PhD Fellowship with grant number 1S52025N and FWO research project PESSO with grant number G018522N.}
}

\author{\IEEEauthorblockN{Mohammad Cheraghinia\IEEEauthorrefmark{1}, Adnan Shahid\IEEEauthorrefmark{1}, Jaron Fontaine\IEEEauthorrefmark{1}, Cedric De Cock\IEEEauthorrefmark{2}, David Plets\IEEEauthorrefmark{2}, Eli De Poorter\IEEEauthorrefmark{1}} \\
\IEEEauthorblockA{\IEEEauthorrefmark{1}\textit{IDLab, Ghent University - imec}, Ghent 9052, Belgium \\
Email: \{mohammad.cheraghinia, adnan.shahid, jaron.fontaine, eli.depoorter\}@ugent.be}
\vspace{0.2cm}
\IEEEauthorblockA{\IEEEauthorrefmark{2}\textit{WAVES, Ghent University - imec}, Ghent 9052, Belgium \\
Email: \{cedric.decock, david.plets\}@ugent.be}
}
\maketitle

\begin{abstract}
User orientation is crucial for many context-aware applications, including interactive museum experiences, smart door access, and intuitive human–environment interaction. However, most existing indoor localization systems focus on estimating position, while body orientation is typically assigned to secondary devices such as inertial measurement units. In this paper, we propose a purely \gls{uwb}-based approach that predicts yaw orientation directly from \gls{uwb} \gls{cir} measurements recorded at fixed anchors as they receive transmissions from a single wearable tag. We use a bidirectional Mamba architecture that captures dependencies across the anchor observations through forward and backward recurrent scans. The model uses per-anchor \gls{cir} and a body-part conditioning module to adapt the representation to different tag placements on the body. Two different Kalman filters are used as post-processing stages to exploit temporal continuity: an orientation-based filter that smooths the neural network predictions, and a location-based filter that additionally incorporates position-derived heading corrections. We evaluated the model's performance in different scenarios to ensure generalizability. The proposed Mamba model achieves a mean absolute error of $38.6^\circ$ in its raw form, outperforming a rule-based baseline $49.5^\circ$. With the location-based Kalman filter, the error is further reduced to $18.9^\circ$, corresponding to a $51\%$ reduction.
\end{abstract}


\section{Introduction}
\label{sec:introduction}

\lettrine{I}{ndoor} localization and tracking have attracted research attention over the past decade for applications in asset management, robotics, and human-computer interaction. \gls{uwb} offers a great solution for high-accuracy indoor positioning due to its fine time resolution, robustness against multipath propagation, and ability to penetrate obstacles~\cite{s16050707}. While \gls{uwb} systems can achieve centimeter-level position accuracy, considering the importance for pedestrian navigation and safety-critical applications, the same \ac{uwb} system is rarely used to estimate human body \emph{orientation} (the direction a person is facing) simultaneously.

Conventional studies on orientation estimation in \gls{uwb} can be divided into three categories. The first category is \emph{geometric multi-tag methods}, which attach two or more \gls{uwb} tags to a body and compute orientation from the relative position vector between them. \cite{johansson2019dual} demonstrated that a dual-tag setup with $30$--$40$\,cm separation can achieve orientation errors below $3^\circ$ under stationary conditions. \cite{zhang2025joint} proposed a cost-effective joint position and orientation estimator using Angle-of-Arrival and Two-Way Ranging with a three-antenna base station; however, the approach requires specialized hardware that is uncommon in wearable deployments.

The second category relies on \emph{\gls{uwb}--inertial fusion}. \cite{s23198289} used IMU heading information to estimate orientation for mitigating human body shadowing in on-body \gls{uwb} tracking, reducing localization errors under non-line-of-sight conditions. More recently, \cite{11213306} presented ObjectTrack, a factor-graph-based 6-DoF tracker that fuses \gls{uwb} ranging with IMU, achieving sub-$10$\,cm position and sub-$5^\circ$orientation errors. \cite{8481661} applied a similar sensor-fusion paradigm to unmanned aerial vehicles, attaining orientation RMSE of $1.93^\circ$. \cite{app15126501} further extended the fusion concept by integrating \gls{uwb}, IMU and visual-inertial odometry within a sliding-window factor graph, reducing position RMSE $67.6\%$ compared to standalone \gls{uwb}. Finally, \cite{10.1145} achieved a median orientation error of $2.7^\circ$ on consumer smartphones by fusing \gls{uwb} angle-of-arrival changes with on-board gyroscope data for short-term drift correction. While these fusion methods are effective, they depend on additional sensors (accelerometers, gyroscopes, or cameras) that increase cost, power consumption, and complexity.

The third category involves \emph{\gls{uwb}-only orientation estimation}. UWBOri \cite{10925144}, calculates device orientation from \gls{uwb} angle and distance measurements. However, this approach relies heavily on multi-antenna, smartphone-grade \gls{uwb} chips. In contrast, our method uses a single-antenna wearable tag that can be a cheaper solution. 

In parallel, deep learning techniques have been increasingly applied to localization tasks. Sequence modeling architectures such as LSTMs~\cite{Tian2024} and Transformers~\cite{info16121033} face trade-offs between temporal modeling capacity and computational efficiency: LSTMs suffer from limited long-range dependency capture, while Transformers have quadratic attention complexity with respect to the sequence length.

Recently, \glspl{ssm} have emerged as an efficient solution for sequence modeling.  Mamba~\cite{gu2024mambalineartimesequencemodeling} extended this method with a selective scan mechanism that allows input-dependent state transitions with linear complexity, which makes it suited for processing UWB \glspl{cir}, which are sequential signals with multipath components. Unlike self-attention mechanisms that scale quadratically with sequence length, Mamba processes long \gls{cir} sequences efficiently, while its input-dependent nature makes the model focus on the most discriminative multipath components and suppress uninformative noise taps. Bidirectional variants of Mamba~\cite{zhang2025lbmambalocallybidirectionalmamba} have demonstrated additional gains. In the \gls{uwb} domain, LightMamba~\cite{11284866} applied a selective \gls{ssm} to \gls{nlos} identification from \gls{cir} data, achieving high accuracy. 
 
In this paper, we propose a bidirectional Mamba-based model for human orientation estimation that uses \gls{uwb} \gls{cir} data from a single tag. Our method requires no additional sensors; unlike geometric methods, it uses a single tag with a single antenna. The model processes per-anchor \gls{cir} vectors and applies bidirectional Mamba across the anchor sequence and predicts yaw orientation. Two Kalman filter variants are introduced for temporal post-processing. Our main contributions are as follows:
\begin{itemize}
    \item We propose a bidirectional Mamba architecture for \gls{uwb}-based orientation estimation that captures directional dependencies in the anchors' \gls{cir} observations with linear computational complexity.
    \item We design a multi-scale convolutional \gls{cir} embedding and a body-part conditioning module that adapts the model to different tag placements.
    \item We introduce orientation-based and location-based Kalman filters as post-processing stages that exploit temporal continuity to smooth predictions.
    \item We evaluate on walking trajectories from three subjects and demonstrate that the proposed method achieves \gls{mae} $18.9^\circ$.
\end{itemize}

This paper is organized as follows. Section~\ref{sec:system_model} formalizes the system model and problem statement. Section~\ref{sec:dataset} describes the experimental setup and dataset. Section~\ref{sec:methodology} presents the proposed model and baselines. Section~\ref{sec:results} reports the experimental results, and Section~\ref{sec:conclusion} concludes the paper.

\section{System Model}
\label{sec:system_model}

We consider an indoor \gls{uwb} positioning environment consisting of $N_{\max}$ fixed anchors deployed at known locations $\{\mathbf{p}^{(n)}\}_{n=1}^{N_{\max}}$ and a single mobile tag carried by a human subject. The tag periodically exchanges ranging messages with the anchors, producing a set of radio measurements at each discrete time step $t$.

\subsection{Measurement Model}

Let $\mathcal{A}_t \subseteq \{1, \dots, N_{\max}\}$ be the subset of anchors that successfully receive the tag’s transmission at time~$t$. For each receiving anchor with index $n \in \mathcal{A}_t$, anchor $n$ records a \gls{cir} vector $\mathbf{c}_t^{(n)}$ of the signal received from the tag, where $C$ is the number of \gls{cir} taps, together with scalar signal-level features such as received signal power and first-path power. The per-anchor observation is denoted:
\begin{equation}
    \mathbf{z}_t^{(n)} = \bigl[\mathbf{c}_t^{(n)\top},\; s_t^{(n)}\bigr]^{\!\top}, \quad n \in \mathcal{A}_t,
    \label{eq:per_anchor_obs}
\end{equation}
where $s_t^{(n)}$ aggregates the scalar features. Because not all anchors are receiving at every time step, a binary availability mask $\mathbf{m}_t \in \{0,1\}^{N_{\max}}$ is maintained, with $m_t^{(n)} = 1$ if $n \in \mathcal{A}_t$ and $0$ otherwise. Observations from unavailable anchors are zero-padded, maintaining a fixed-size multi-anchor measurement at time~$t$:
\begin{equation}
    \mathbf{Z}_t = \bigl\{\mathbf{z}_t^{(n)} \cdot m_t^{(n)}\bigr\}_{n=1}^{N_{\max}}.
    \label{eq:multi_anchor_obs}
\end{equation}

The \gls{cir} includes the information of the propagation channel. When the subject rotates, the human body attenuates signals from anchors, resulting in orientation-based variations.

\subsection{Problem Formulation}

Over a measurement session of $T$ time steps, the system produces a temporal observation sequence
\begin{equation}
    \mathcal{X} = \bigl\{(\mathbf{Z}_t,\; \mathbf{m}_t,\; b)\bigr\}_{t=1}^{T},
    \label{eq:obs_sequence}
\end{equation}
where $b$ is a categorical variable indicating the body-part placement of the tag (e.g., chest, pocket, or arm).

Let $\theta_t \in (-180^{\circ}, 180^{\circ}]$ denote the ground-truth yaw angle of the subject at time~$t$. The goal is to learn a mapping
\begin{equation}
    f_\Theta : \mathcal{X} \;\longrightarrow\; \{\hat{\theta}_t\}_{t=1}^{T},
    \label{eq:problem}
\end{equation}

\begin{figure}[t]
    \centering
    \begin{subfigure}{0.48\columnwidth}
        \centering
        \includegraphics[width=0.9\linewidth]{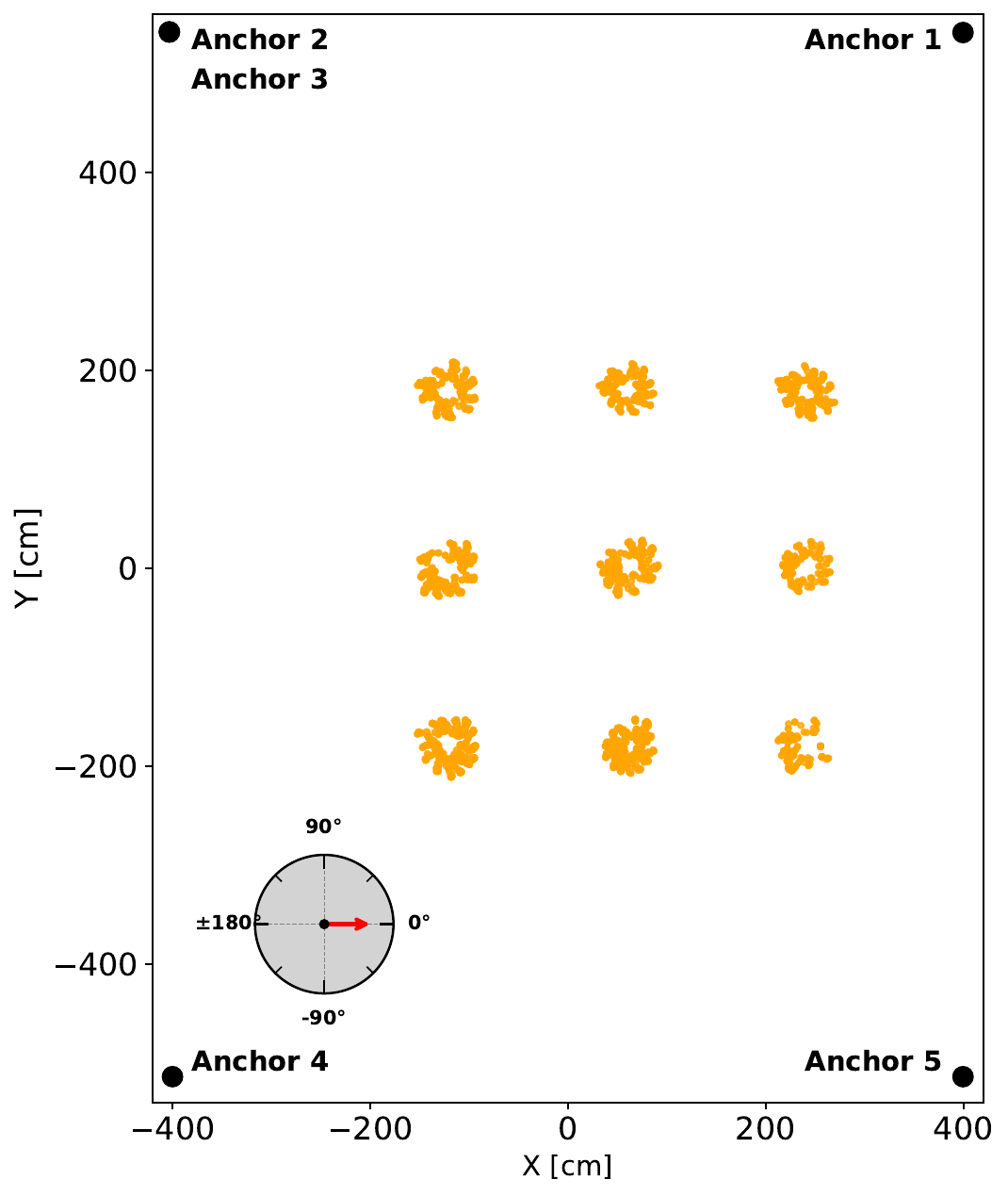}
        \caption{Training and validation samples showing rotating in place.}
        \label{fig:image1}
    \end{subfigure}
    \hfill 
    \begin{subfigure}{0.48\columnwidth}
        \centering
        \includegraphics[width=0.9\linewidth]{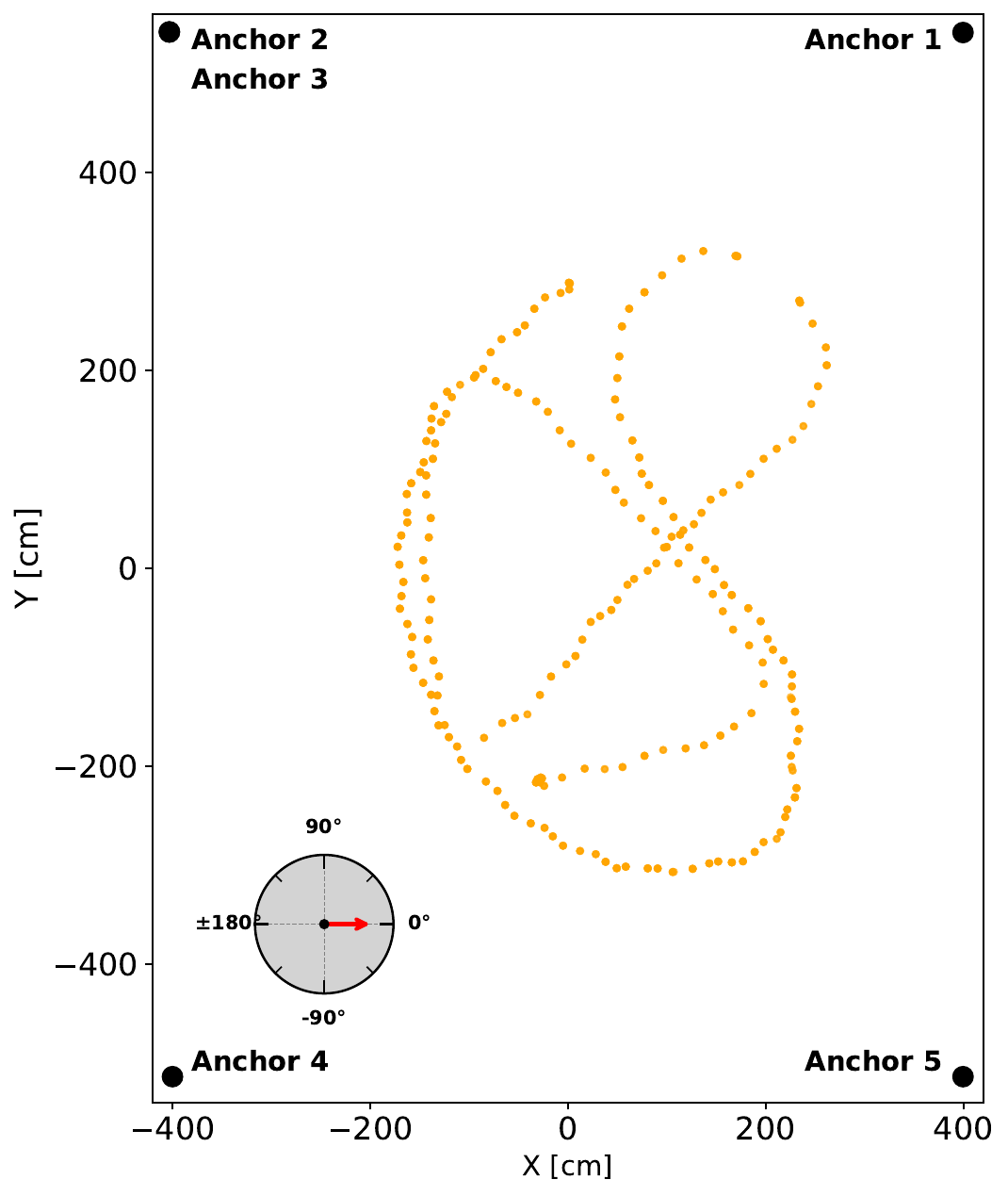}
        \caption{Testing samples showing walking in a specific trajectory.}
        \label{fig:image2}
    \end{subfigure}
    
    \caption{Data samples for one subject. Black dots are the five anchors, and orange dots show the \gls{mocap} locations.}
    \label{fig:combined_figures}
\end{figure}

\noindent that estimates the yaw angle from the multi-anchor \gls{uwb} observations, where $\Theta$ denotes the set of learnable parameters. Because the yaw angle is a periodic quantity with the domain wrapping at $\pm 180^{\circ}$, we adopt a continuous angular representation by predicting $(\cos\hat{\theta}_t,\, \sin\hat{\theta}_t)$ and recovering the angle via $\hat{\theta}_t = \operatorname{atan2}(\sin\hat{\theta}_t,\, \cos\hat{\theta}_t)$, avoiding discontinuities at the boundary.

The mapping $f_\Theta$ is implemented by a deep sequence model that operates on the ordered anchor observations at each time step (Section~\ref{sec:methodology}), while Kalman-filter post-processing leverages temporal continuity across time steps.

\section{Experimental Setup and Dataset}
\label{sec:dataset}
\subsection{Experimental Setup}

Data collection was carried out in the IDLab Industrial IoT~(IIoT) Lab at Ghent University \cite{idlab_iiot_lab}, an indoor testbed equipped with \gls{uwb} infrastructure and a high-precision \gls{mocap} system. Five DW1000-based \gls{uwb} anchors were deployed throughout the environment at known positions \cite{s19071548}, as indicated by the black dots in Fig.~\ref{fig:combined_figures}. A single \gls{uwb} tag, also based on the DW1000 transceiver, was carried by the subject and communicated with the anchors at an approximate rate of $2.2$\,Hz. At each time step (superframe), each visible anchor recorded a \gls{cir} vector of the signal received from the tag, along with the received signal power and first-path power. The \gls{mocap} system simultaneously captured the subject's position and full $3 \times 3$ rotation matrix at each superframe, from which the ground-truth yaw angle was extracted. The \gls{mocap} positions are shown as orange dots in Fig.~\ref{fig:combined_figures}.

\subsection{Dataset and Subjects}

Three participants took part in the data collection, each wearing the \gls{uwb} tag at three body locations: the chest, the pocket, and the arm. These placements are common wearable positions and exhibit distinct signal attenuation and \gls{cir} characteristics; for instance, chest-mounted tags produce a symmetric body-shadowing pattern, whereas arm-mounted tags introduce directional variation due to limb motion.

The dataset includes two types of movement. For training and validation, performed in-place rotations at nine fixed locations (Fig.~\ref{fig:image1}), ensuring full $360^{\circ}$ orientation coverage with diverse \gls{cir} patterns. An $80$/$20$ random split is used to separate the training and validation sets. For testing, subjects walked in predefined trajectories (Fig.~\ref{fig:image2}), to evaluate the models' generalization from rotations to realistic motion where both position and orientation change simultaneously.

\section{Methodology}
In this section, we discuss our proposed solutions. 

\label{sec:methodology}

\subsection{Proposed UWB Bidirectional Mamba Model}
\label{subsec:bimamba}

\begin{figure}[t]
    \centering
    \includegraphics[width=0.9\linewidth]{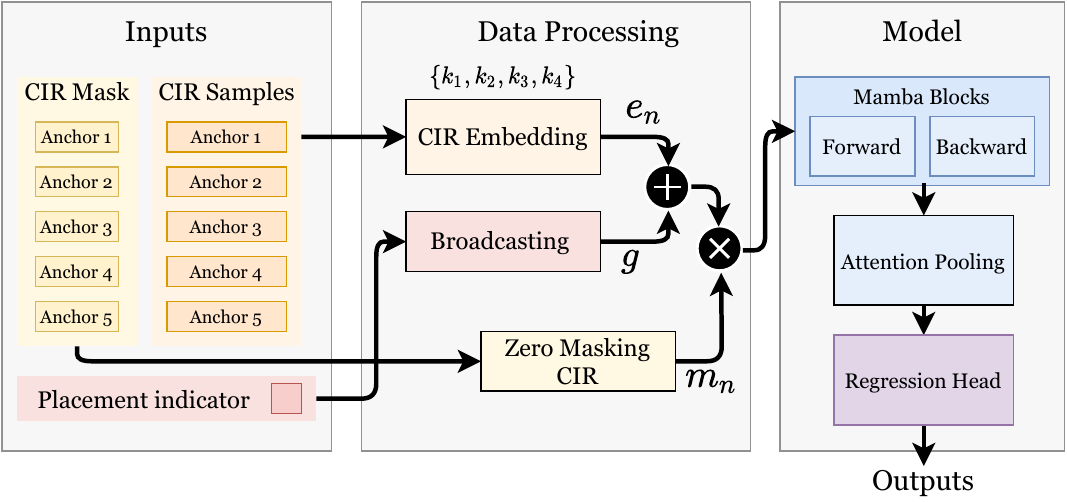}
    \caption{Architecture of the proposed Mamba‑based solution.}
    \label{fig:mamba_arch}
\end{figure}

The proposed model takes as input the \gls{cir} samples collected from all available \gls{uwb} anchors, a binary mask indicating anchor availability, and an optional auxiliary input (e.g., a body-part placement indicator). The architecture, illustrated in Fig.~\ref{fig:mamba_arch}, consists of two stages: (1) data processing and (2) a model followed by a regression head.

The input at each time step follows the measurement model of Section~\ref{sec:system_model}: per-anchor \gls{cir} vectors and the binary availability mask are provided as defined in Eqs.~\eqref{eq:per_anchor_obs}--\eqref{eq:multi_anchor_obs}.

Each per-anchor \gls{cir} vector is processed by a shared multi-scale convolutional feature extractor that maps the raw $C$-tap signal into a compact $d$-dimensional embedding. We treat the \gls{cir} as a one-dimensional time series rather than a flat feature vector because the orientation-relevant information is encoded in the local shape of the multipath profile: the leading-edge slope, the first-path-to-peak spacing, and the relative attenuation of later taps all vary as the human body shadows the direct path.

To capture multipath structure at several temporal scales simultaneously, we use four parallel branches with kernel sizes $\{3, 5, 7, 15\}$. Narrow kernels resolve sharp, sample-level features such as the first-path onset, whereas wider kernels capture the broader delay-spread. The kernel sizes are derived from the \gls{cir} length so that the receptive fields scale with the signal resolution. Each branch consists of two convolutional layers, each followed by batch normalization and ReLU activation, producing the branch output $\mathbf{f}_n^{(i)}$ for $i = 1,\dots,4$. Stacking two layers enlarges the effective receptive field and adds non-linearity, while batch normalization stabilizes training across the wide dynamic range of \gls{cir} magnitudes observed at different anchor distances. Each branch is assigned $d/4$ channels so that the four scales share the embedding budget equally and concatenate to exactly $d$ channels. An adaptive average pooling layer then reduces each branch output to a fixed length, which decouples the embedding dimension from the \gls{cir} length and yields a representation of constant size regardless of $C$. We use average rather than max pooling because the body-shadowing cue manifests as a distributed reduction in tap energy, which average pooling preserves, rather than as a single dominant peak. The four pooled feature maps, each of dimension $d/4$ channels, are concatenated along the channel axis and linearly projected to yield the per-anchor embedding:

\begin{equation}
    \mathbf{e}_n = \mathbf{W}_{\mathrm{proj}} \mathop{\|}_{i=1}^4 [\text{AvgPool}(\mathbf{f}_n^{(i)})] \in \mathbb{R}^{d},
\end{equation}
where $d$ is the embedding dimension and $\mathbf{W}_{\mathrm{proj}}$ is a learnable projection matrix that fuses the multi-scale features into a single embedding. The same extractor weights are shared across all anchors, so the network learns one generic \gls{cir}-to-feature mapping rather than an anchor-specific one; this keeps the parameter count independent of $N_{\max}$ and makes the embedding invariant to which anchors happen to be available at a given time step.

The body-shadowing of a rotation depends strongly on where the tag is worn. To let a single model account for the differences, we provide the placement as an explicit conditioning signal instead of training a separate model per body part. The body-part placement indicator $b$ from Eq.~\eqref{eq:obs_sequence} is mapped to the embedding space via a two-layer conditioner:
\begin{equation}
    \mathbf{g} = \mathbf{W}_2 \text{GELU}(\text{LN}(\mathbf{W}_1 \mathbf{b})) \in \mathbb{R}^{d},
\end{equation}
where $\mathbf{W}_1$, $\mathbf{W}_2$ are learnable parameters, and $\text{LN}(\cdot)$ denotes layer normalization. The conditioning vector is added element-wise to every anchor embedding, a lightweight additive (FiLM-style) scheme that shifts the shared representation into a placement-specific region of the embedding space without altering the downstream architecture or its parameter count. At the end, zero masking is applied element-wise to hide the embeddings from anchors that are unavailable.
\begin{equation}
    \tilde{\mathbf{e}}_n = (\mathbf{e}_n + \mathbf{g}) \cdot m_n, \quad n = 1, \dots, N_{\max}.
\end{equation}

The masked embedding sequence $\mathbf{X} = [\tilde{\mathbf{e}}_1, \dots, \tilde{\mathbf{e}}_{N_{\max}}] \in \mathbb{R}^{N_{\max} \times d}$ is processed by  $L$ bidirectional Mamba layers.

A single Mamba block implements a selective \gls{ssm}. Given an input sequence $\mathbf{X}$, the block first projects it into two branches via a linear projection:
\begin{equation}
    [\mathbf{X}', \mathbf{Z}] = \text{split}(\mathbf{W}_{\mathrm{in}} \, \mathbf{X}), \quad \mathbf{X}', \mathbf{Z} \in \mathbb{R}^{N_{\max} \times d_{\mathrm{in}}},
\end{equation}
where $d_{\mathrm{in}} = \alpha \cdot d$ with expansion factor $\alpha$ and $\mathbf{W}_{\mathrm{in}} \in \mathbb{R}^{2d_{\mathrm{in}} \times d}$. The branch $\mathbf{X}'$ is passed through a depthwise 1-D convolution, followed by a SiLU activation with an output of $\hat{\mathbf{X}}$.

Following the selective state space formulation of Mamba~\cite{gu2024mambalineartimesequencemodeling}, the activated feature $\hat{\mathbf{X}}$ is used to generate input-dependent \gls{ssm} parameters $\boldsymbol{\Delta}_t$, $\mathbf{B}_t$, and $\mathbf{C}_t$ at each time step via learned projections, where $N$ denotes the \gls{ssm} state dimension. These are discretized and applied through a recurrent scan:
\begin{align}
    \mathbf{h}_t &= \bar{\mathbf{A}}_t \odot \mathbf{h}_{t-1} + \bar{\mathbf{B}}_t \odot \hat{\mathbf{x}}_t^{\top}, \quad \mathbf{h}_0 = \mathbf{0}, \label{eq:state_update} \\
    \mathbf{y}_t &= \mathbf{h}_t \, \mathbf{C}_t + \hat{\mathbf{x}}_t \odot \mathbf{D}, \label{eq:ssm_output}
\end{align}
where $\bar{\mathbf{A}}_t$ and $\bar{\mathbf{B}}_t$ are the discretized state and input matrices, $\mathbf{D} \in \mathbb{R}^{d_{\mathrm{in}}}$ is a learnable skip-connection parameter, and $\odot$ denotes the Hadamard product.

The \gls{ssm} output is combined with the second branch via element-wise gating and projected back to the model dimension:
\begin{equation}
    \mathbf{O} = \mathbf{W}_{\mathrm{out}}\bigl(\mathbf{Y} \odot \operatorname{SiLU}(\mathbf{Z})\bigr), \quad \mathbf{W}_{\mathrm{out}} \in \mathbb{R}^{d \times d_{\mathrm{in}}}.
\end{equation}

Each of the $L$ layers applies a forward and a backward Mamba block with normalization and residual connections. The backward branch operates on the sequence-reversed input. The two directional outputs are concatenated and fused:
\begin{equation}
    \mathbf{X}_\ell = \mathbf{W}_{\mathrm{fuse}}\, \operatorname{concat}\!\bigl[\mathbf{X}^{\rightarrow}_\ell,\, \mathbf{X}^{\leftarrow}_\ell\bigr], \quad \mathbf{W}_{\mathrm{fuse}} \in \mathbb{R}^{d \times 2d},
\end{equation}
where $\ell = 1, \dots, L$. After fusion, the mask is re-applied to zero out unavailable anchors.

A learnable query $\mathbf{q}$ aggregates the sequence via multi-head cross-attention with mask. The pooled vector is passed through a regression head with an $\ell_2$-normalized pair from which the yaw is recovered via $\operatorname{atan2}$.

\subsection{Kalman Filters}
\label{subsec:kalman}

For human tracking, often frequent positioning updates are provided, allowing the use of temporal information in addition to the raw orientation estimation from Mamba. We apply Kalman filters as a post-processing stage to smooth the Mamba outputs over time. Both variants share a two-dimensional state $\mathbf{x}_k = [\theta_k,\, \dot{\theta}_k]^\top$ (yaw angle and yaw rate) with a constant yaw-rate transition model. In tracking literature this is also referred to as a random-walk model on the yaw rate, since the process noise on $\dot{\theta}$ allows the filter to progressively forget its prior yaw-rate estimate and adapt to changing turn dynamics~\cite{bar2004estimation}. The prediction step propagates the state and covariance as:
\begin{align}
    \hat{\mathbf{x}}_{k|k-1} &= \mathbf{F}_k \, \mathbf{x}_{k-1|k-1}, \quad
    \mathbf{F}_k = \begin{bmatrix} 1 & \Delta t_k \\ 0 & 1 \end{bmatrix}, \label{eq:kf_state_pred} \\
    \mathbf{P}_{k|k-1} &= \mathbf{F}_k \, \mathbf{P}_{k-1|k-1} \, \mathbf{F}_k^\top + \mathbf{Q}_k, \label{eq:kf_cov_pred}
\end{align}
\begin{algorithm}[t]
\caption{\gls{lut}-Based Yaw Prediction}
\label{alg:yaw_prediction_lut}
\begin{algorithmic}[1]
\Require $\mathcal{F} = \{(a, \text{rxp}, y)\}$, scenario $s \in \{1, 2\}$, \gls{lut} $\mathcal{T}$
\Ensure Predicted yaw $\hat{y}$, actual yaw $y$
\State $y \leftarrow$ \text{Ground-truth yaw}
\State $rxp \leftarrow$ \text{Received Power}
\State $a \leftarrow$ \text{Anchor ID}

\State Sort $\mathcal{F}$ by $\text{rxp}$ descending

\State $a_1 \leftarrow$ \text{Strongest rxp anchor ID}
\State $a_2 \leftarrow$ \text{Second strongest rxp anchor ID}

\State $(lo, hi) \leftarrow \mathcal{T}[a_1, a_2]$ \text{retrieval from \gls{lut}}

\State \textbf{return} $\text{GetAngle}(lo, hi, s), y$

\Function{GetAngle}{$lo, hi, s$}
    \If{$s = 1$}
        \State \Return $\mathcal{U}(lo, hi)$ \text{(Uniform random sample)}
    \ElsIf{$s = 2$}
        \State \Return $(lo + hi)\;/\;2$ \text{(Bin center)}
    \EndIf
\EndFunction
\end{algorithmic}
\end{algorithm}

\begin{table}[t]
    \centering
    \caption{Orientation Prediction Lookup Table (\gls{lut})($\mathcal{T}$)}
    \label{tab:yaw_lut}
    \begin{tabular}{llr}
        \toprule
        \textbf{Strongest} ($a_1$) & \textbf{Second Strongest} ($a_2$) & \textbf{Range} $(lo, hi)$ \\
        \midrule
        Anchor1 & Anchor2 or 3 & $(-45^\circ, 0^\circ)$ \\
        Anchor1 & Anchor5 & $(-90^\circ, -45^\circ)$ \\
        \midrule
        Anchor2 or 3 & Anchor1 & $(0^\circ, 45^\circ)$ \\
        Anchor2 or 3 & Anchor4 & $(45^\circ, 90^\circ)$ \\
        \midrule
        Anchor4 & Anchor2 or 3 & $(90^\circ, 135^\circ)$ \\
        Anchor4 & Anchor5 & $(135^\circ, 180^\circ)$ \\
        \midrule
        Anchor5 & Anchor1 & $(-135^\circ, -90^\circ)$ \\
        Anchor5 & Anchor4 & $(-180^\circ, -135^\circ)$ \\
        \bottomrule
    \end{tabular}
\end{table} 
where $\Delta t_k$ is the time step between measurements $k{-}1$ and $k$, and $\hat{\mathbf{x}}_{k|k-1}$ denotes the predicted state. The time-dependent process noise covariance $\mathbf{Q}_k$ is constructed from the yaw noise variance $\sigma^2_\theta$ and the yaw-rate noise variance $\sigma^2_{\dot{\theta}}$ as:
\begin{equation}
    \mathbf{Q}_k = \begin{bmatrix} \sigma^2_\theta \Delta t_k + \frac{1}{3}\sigma^2_{\dot{\theta}} \Delta t_k^3 & \frac{1}{2}\sigma^2_{\dot{\theta}} \Delta t_k^2 \\ \frac{1}{2}\sigma^2_{\dot{\theta}} \Delta t_k^2 & \sigma^2_{\dot{\theta}} \Delta t_k \end{bmatrix}, \label{eq:kf_Q}
\end{equation}
which corresponds to a piecewise-constant white-noise jerk model integrated over $\Delta t_k$, plus an additive yaw diffusion term $\sigma^2_\theta$. This formulation accounts for the variable time intervals produced by the \gls{uwb} superframe schedule. Following standard Kalman filter notation, $\mathbf{P}_{k|k-1}$ and $\mathbf{P}_{k|k}$ denote the predicted and updated error covariance matrices, $\mathbf{K}_k$ is the Kalman gain, and $\mathbf{I}_2$ is the $2{\times}2$ identity matrix. The operator $\operatorname{wrap}(\cdot)$ maps angles to $[-180^{\circ}, 180^{\circ}]$ and $\operatorname{clip}(x, a, b)$ clamps $x$ to $[a, b]$.

\subsubsection{Orientation-based Kalman Filter}
\label{subsubsec:kf_yaw}

This filter uses only the neural network yaw prediction $z_k^{\mathrm{nn}}$ as the measurement, making it suitable even in conditions where the current position is not known to the tag. The measurement model is $\mathbf{H} = [1,\, 0]$, and the standard Kalman update is applied with measurement noise variance $R^{\mathrm{nn}}$:
\begin{align}
    \tilde{y}_k &= \operatorname{wrap}\!\bigl(z_k^{\mathrm{nn}} - \hat{\theta}_{k|k-1}\bigr), \\
    \mathbf{K}_k &= \mathbf{P}_{k|k-1} \, \mathbf{H}^\top \bigl(\mathbf{H} \, \mathbf{P}_{k|k-1} \, \mathbf{H}^\top + R^{\mathrm{nn}}\bigr)^{-1}, \\
    \mathbf{x}_{k|k} &= \hat{\mathbf{x}}_{k|k-1} + \mathbf{K}_k \, \tilde{y}_k, \quad
    \mathbf{P}_{k|k} = (\mathbf{I}_2 - \mathbf{K}_k \, \mathbf{H}) \, \mathbf{P}_{k|k-1}.
\end{align}

\subsubsection{Location-based Kalman Filter}
\label{subsubsec:kf_location}

\begin{figure}[t]
    \centering
    \includegraphics[width=0.9\linewidth]{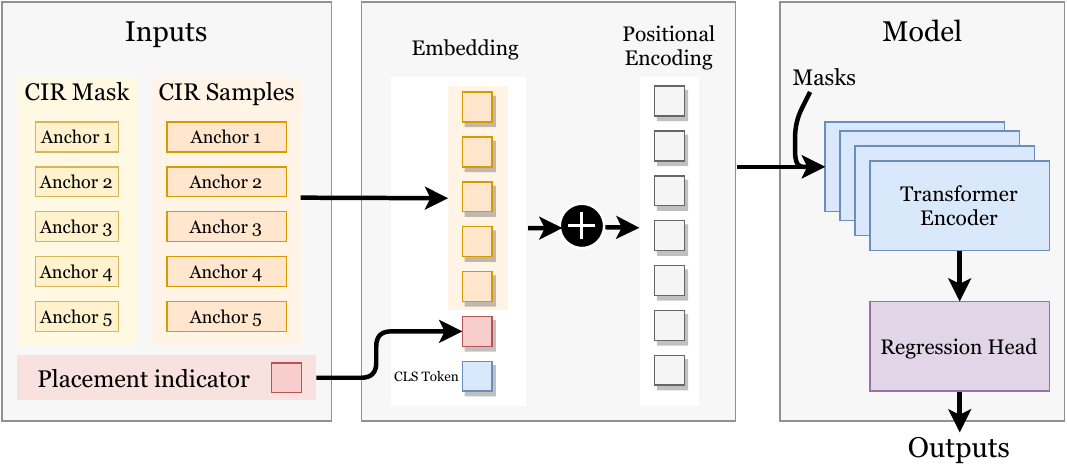}
    \caption{Architecture of the transformer‑based baseline.}
    \label{fig:transformer}
\end{figure}

This filter extends the orientation-based variant by incorporating a heading measurement derived from the tag's own movement between consecutive time steps. After the neural network update (identical to above, yielding $\mathbf{x}_k^{(1)}$, $\mathbf{P}_k^{(1)}$), the Euclidean displacement of the tag $d_k = \|\mathbf{p}_k - \mathbf{p}_{k-1}\|$ is computed from two successive positions $\mathbf{p}_k = (x_k, y_k)$. When the tag has moved more than a minimum threshold $d_{\min}$, the direction of that movement is used as a second orientation measurement:
\begin{equation}
    z_k^{\mathrm{pos}} = \operatorname{atan2}(\Delta y_k,\, \Delta x_k), \quad
    R_k^{\mathrm{pos}} = \frac{R^{\mathrm{pos}}_0}{\operatorname{clip}(10\,d_k,\, 1,\, 5)},
\end{equation}
where the noise variance is scaled inversely with displacement magnitude. A second Kalman update is then applied to $\mathbf{x}_k^{(1)}$, $\mathbf{P}_k^{(1)}$ using $z_k^{\mathrm{pos}}$ and $R_k^{\mathrm{pos}}$. When $d_k \leq d_{\min}$, the second update is skipped.
\begin{table*}[htbp]
    \centering
    \footnotesize
    \caption{Comparison of model architectures, including modality, temporal processing, location awareness, update rate, and corresponding \ac{mae} performance.}
    \label{tab:combined_method_model_comparison}
    \setlength{\tabcolsep}{4pt}
    \renewcommand{\arraystretch}{1}
    \begin{tabular}{@{} 
       | >{\raggedright\arraybackslash}m{3.7cm} |   
        >{\centering\arraybackslash}m{1cm} | 
        >{\centering\arraybackslash}m{1.3cm}   |   
        >{\centering\arraybackslash}m{2.2cm} | 
        >{\centering\arraybackslash}m{1.7cm}   | 
        >{\centering\arraybackslash}m{1.5cm} | 
        >{\centering\arraybackslash}m{0.7cm}     
        >{\centering\arraybackslash}m{0.7cm}     
        >{\centering\arraybackslash}m{0.7cm}     
        >{\centering\arraybackslash}m{0.9cm}  |   
        @{}}
        \toprule

        \textbf{Method} & 
        \textbf{Modality} & 
        \textbf{Body-part} & 
        \textbf{Temporal process} & 
        \textbf{Location info} & 
        \textbf{Update rate} & 

        \multicolumn{4}{c}{\textbf{\gls{mae} ($^\circ$) $\downarrow$}} \\
        \cmidrule(l){7-10}
        
        & & & & & & \textbf{Chest} & \textbf{Pocket} & \textbf{Arm} & \textbf{TOTAL} \\
        \midrule
        
         Rule-based (S1) & Power & $\times$ & $\times$ & $\times$ & Low  & 42.9 & 47.6 & 58.1 & 50.5 \\
         Rule-based (S2) & Power & $\times$ & $\times$ & $\times$ & Low  & 41.3 & 47.3 & 57.2 & 49.5 \\
         MoCap Heading & Optical & $\times$ & $\times$ & \checkmark & High & 88.2 & 77.5 & 27.2 & 59.6 \\
        \midrule

         Transformer (raw) & CIR & \checkmark & $\times$ & $\times$ & Low  & 50.3 & 46.1 & 40.8 & 45.2 \\
        Transformer + KF (orientation) & CIR & \checkmark & \checkmark & $\times$ & High  & 46.6 & 39.5 & 36.4 & 40.5 \\
         Transformer + KF (loc.-based) & CIR & \checkmark & \checkmark & \checkmark & High  & 25.6 & 13.1 & 17.7 & 19.3 \\
        \midrule

         Mamba (raw) & CIR & \checkmark & $\times$ & $\times$ & Low  & 40.5 & 41.3 & 35.8 & 38.6 \\
        Mamba + KF (orientation) & CIR & \checkmark & \checkmark & $\times$ & High & 37.9 & 32.0 & 30.7 & 33.4 \\
        Mamba + KF (location) & CIR & \checkmark & \checkmark & \checkmark & High  & 24.1 & 13.0 & 17.9 & 18.9 \\
        \bottomrule
        
    \end{tabular}
\end{table*}

In the current implementation, the positions $\mathbf{p}_k$ are obtained from the high-accuracy \gls{mocap} system. This choice provides an upper bound on the location-based filter's capability but these values can be replaced by \gls{uwb}-derived locations.

\subsection{Training Details} \label{subsec:training}
Mamba and Transformer models are trained to minimize the \ac{mse}. We use the AdamW optimizer with a learning rate of $10^{-4}$ and a batch size of $32$, training for $200$ epochs. Each \gls{cir} vector consists of $C = 150$ samples. The embedding dimension is $d = 64$, with $L = 4$ layers, $4$ attention heads, and a dropout rate of $0.165$. The Kalman filter parameters were tuned via Bayesian optimization.

\begin{figure}[t]
    \centering
    \includegraphics[width=0.9\linewidth]{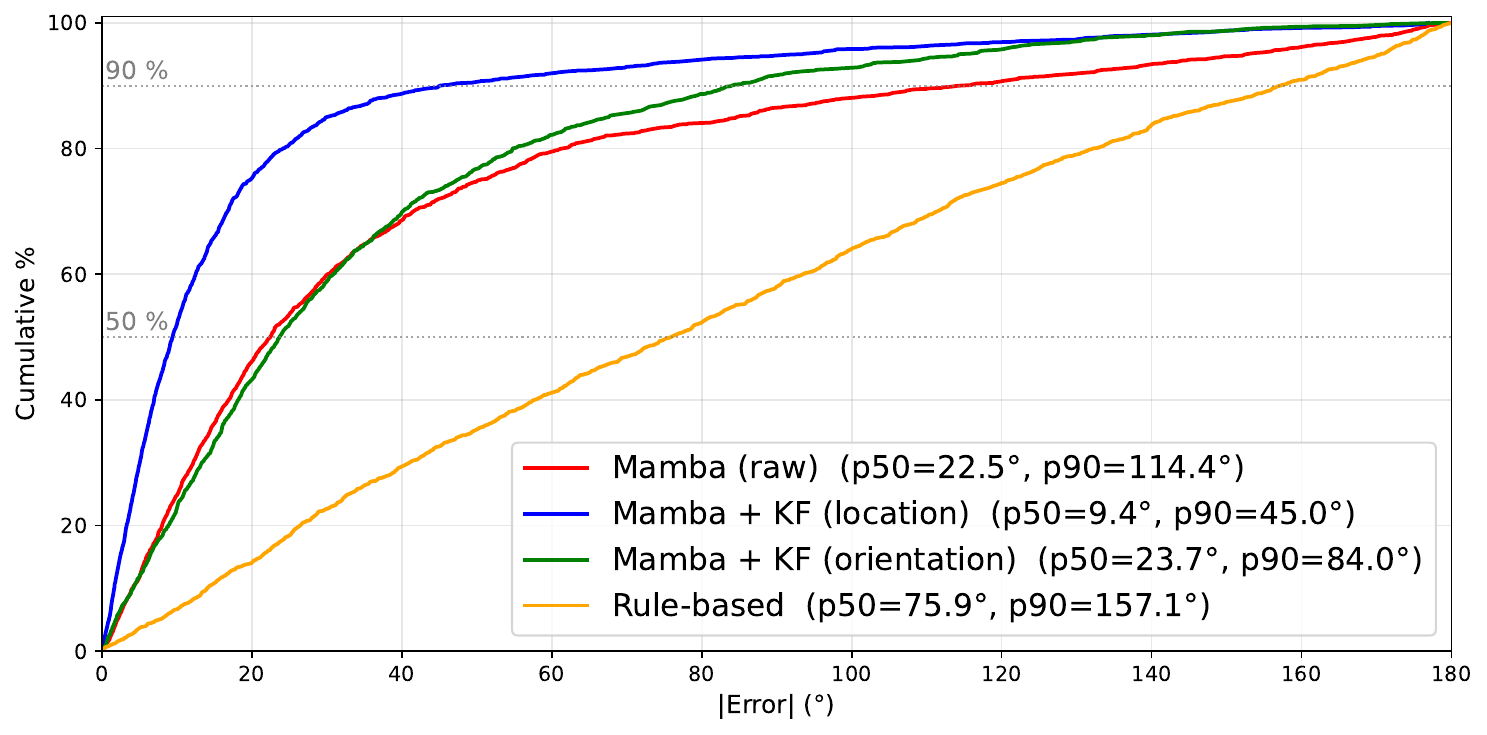}
        \caption{CDF of yaw errors across all test sequences for the Mamba model with Median~(p50) and 90th~percentile~(p90).}

    \label{fig:cdf}
\end{figure} 

\section{Results}
\label{sec:results}

We evaluate the proposed model against three baselines: (i) a rule-based Look-Up Table (\gls{lut})-based yaw prediction method (summarized in Algorithm~\ref{alg:yaw_prediction_lut} and Table~\ref{tab:yaw_lut}); (ii) a \gls{mocap}-based heading (\gls{mocap} Hdg), computed from ground-truth position estimates as a reference that quantifies the achievable position-derived heading alone \textbf{without knowing where the tag is attached}; and (iii) a Transformer-based deep learning model that captures dependencies via self-attention (Fig.~\ref{fig:transformer}). For our Transformer model, three output variants are reported: the raw network output, the orientation Kalman filter, and the location Kalman filter.
The models are evaluated on walking test sequences collected from three subjects across three body-part placements (Chest, Pocket, Arm). The \gls{mae} is the metric of the error in range of $[-180^{\circ}, 180^{\circ}]$.

The architectural characteristics and corresponding \gls{mae} results for all methods are summarized in Table~\ref{tab:combined_method_model_comparison}. The rule-based approach, which operates without temporal state and requires only received signal power, can be a compute-constrained fallback; its angular binning limits accuracy to $49.5^\circ$ the best variant~(S2). The raw Mamba, similarly is state-independent but processing per anchor \gls{cir} with body-part conditioning, achieves $38.6^\circ$, outperforming both the rule-based baselines and the Transformer~($45.2^\circ$), without needing any position information, \textbf{making it the most reliable solution for a single tag}. Introducing temporal processing via orientation Kalman filters reduces Mamba's error to $33.4^\circ$, using continuity between predictions, and remains the choice for orientation estimation when position is unavailable. Location Kalman filter mixes position-based heading corrections at a high update rate, achieving $18.9^\circ$ with a $51\%$ reduction relative to the raw Mamba at the cost of requiring a position input. The \gls{mocap} Hdg, which relies solely on ground-truth positions without any knowledge of tag placement, yields $59.6^\circ$, demonstrating that position information alone is insufficient when placement context is absent. The Transformer with location Kalman filter ($19.3^\circ$) is competitive, but Mamba maintains a clear advantage in any configuration.

\begin{figure*}[t]
    \centering
    \begin{subfigure}[b]{0.18\textwidth}
        \centering
        \includegraphics[width=\textwidth]{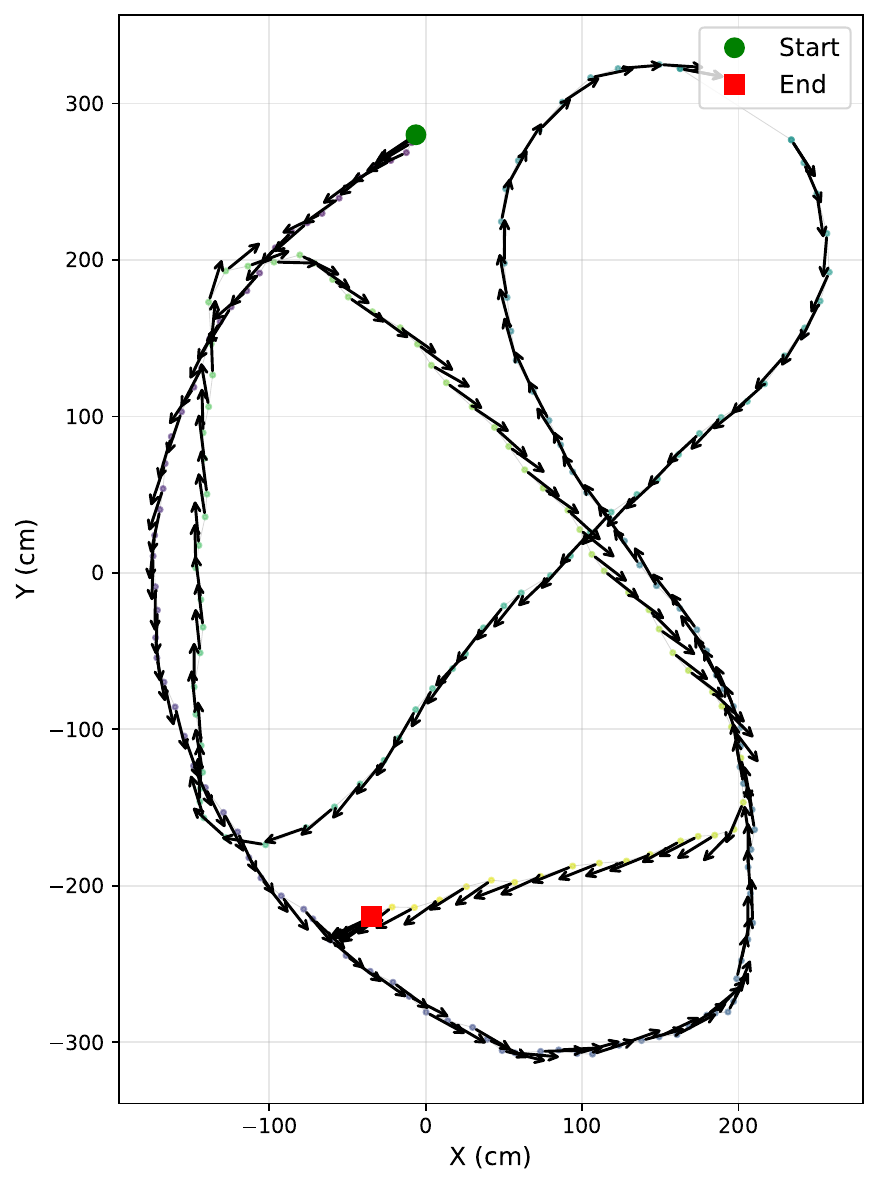}
        \caption{Ground Truth}
        \label{fig:ground_truth}
    \end{subfigure}
    \hfill
    \begin{subfigure}[b]{0.18\textwidth}
        \centering
        \includegraphics[width=\textwidth]{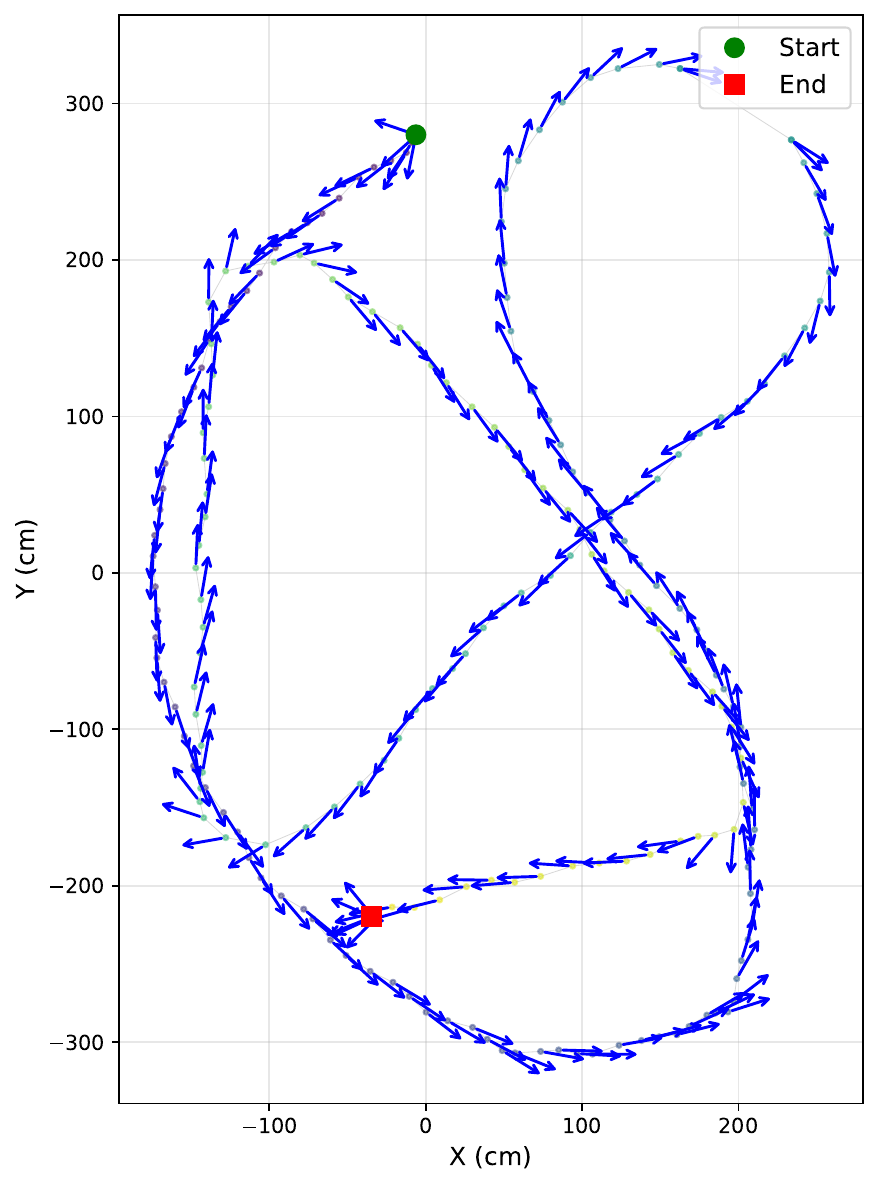}
        \caption{Location KF}
        \label{fig:kf_location}
    \end{subfigure}
    \hfill
    \begin{subfigure}[b]{0.18\textwidth}
        \centering
        \includegraphics[width=\textwidth]{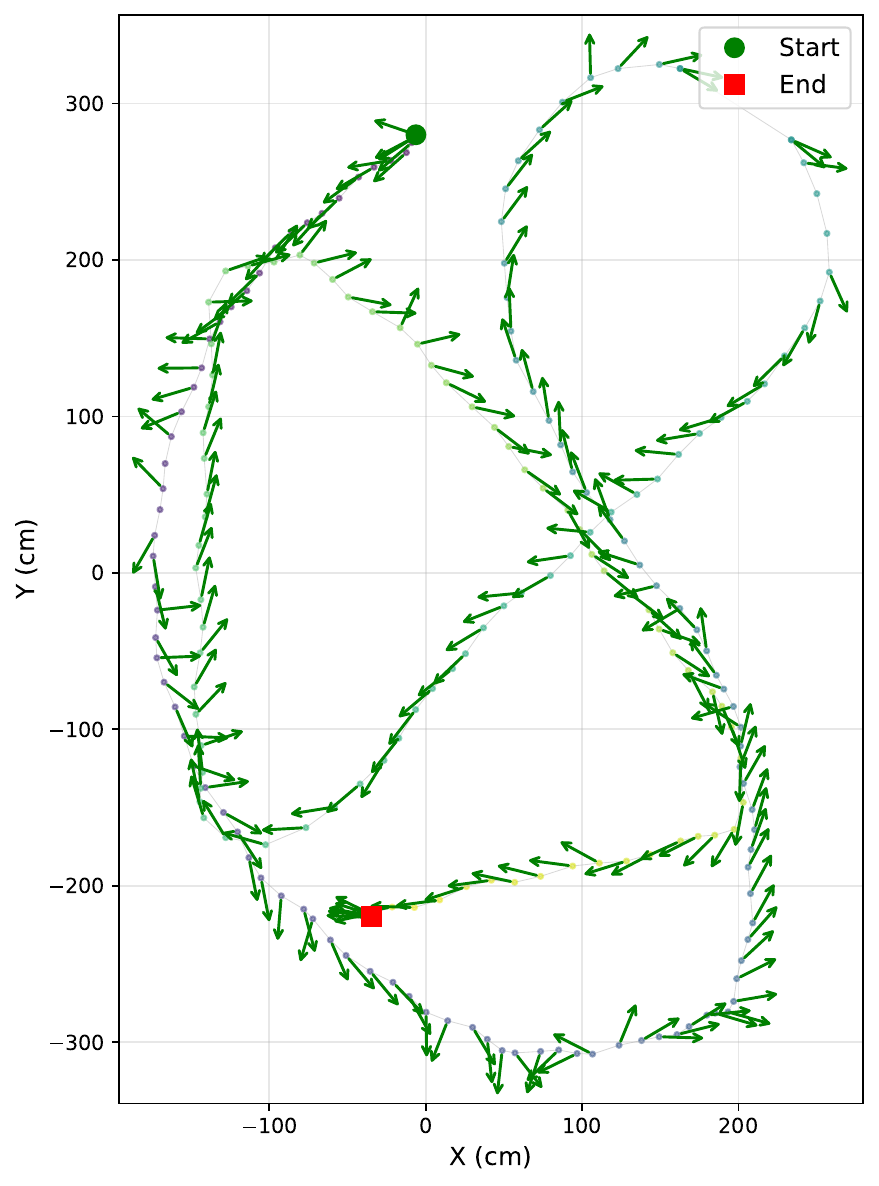}
        \caption{Orientation KF}
        \label{fig:kf_yaw}
    \end{subfigure}
    \hfill
    \begin{subfigure}[b]{0.18\textwidth}
        \centering
        \includegraphics[width=\textwidth]{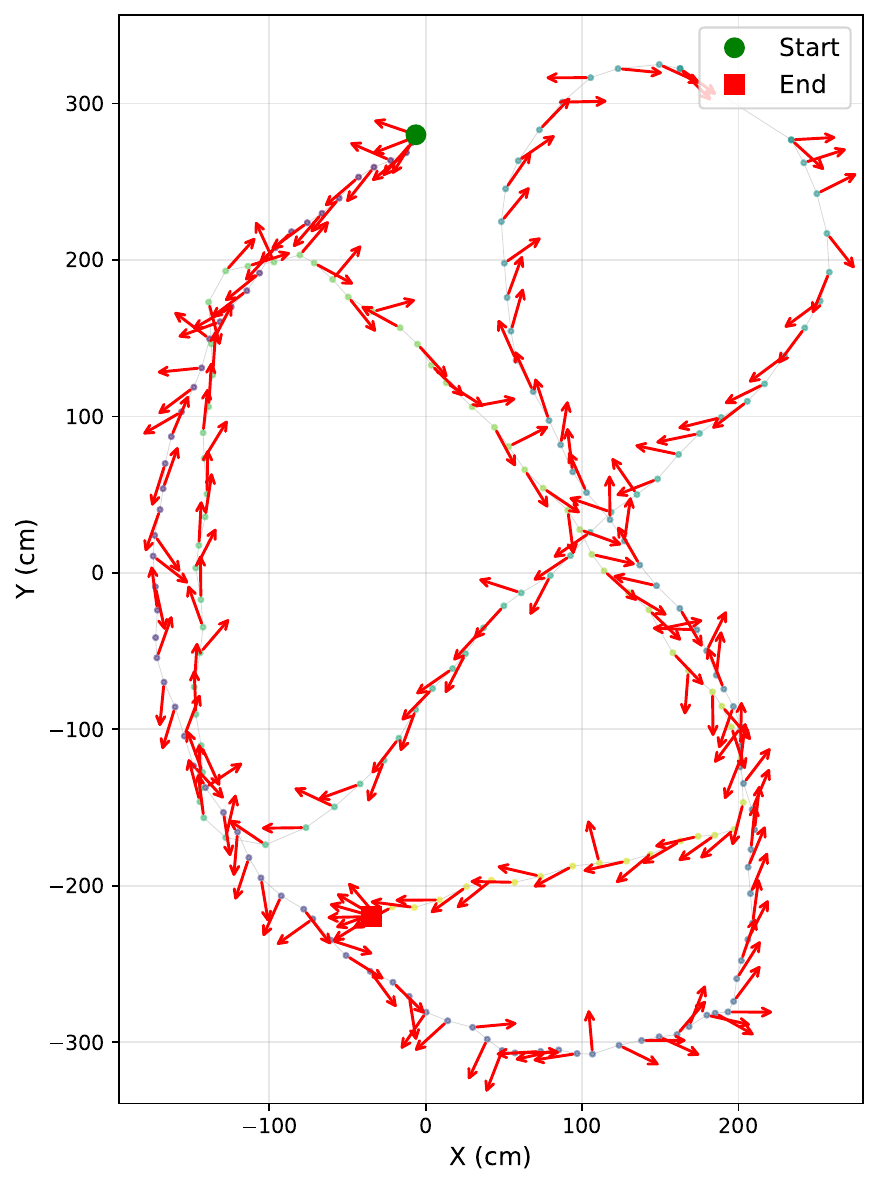}
        \caption{Mamba Raw}
        \label{fig:raw_nn}
    \end{subfigure}
    \hfill 
    \begin{subfigure}[b]{0.18\textwidth}
        \centering
        \includegraphics[width=\textwidth]{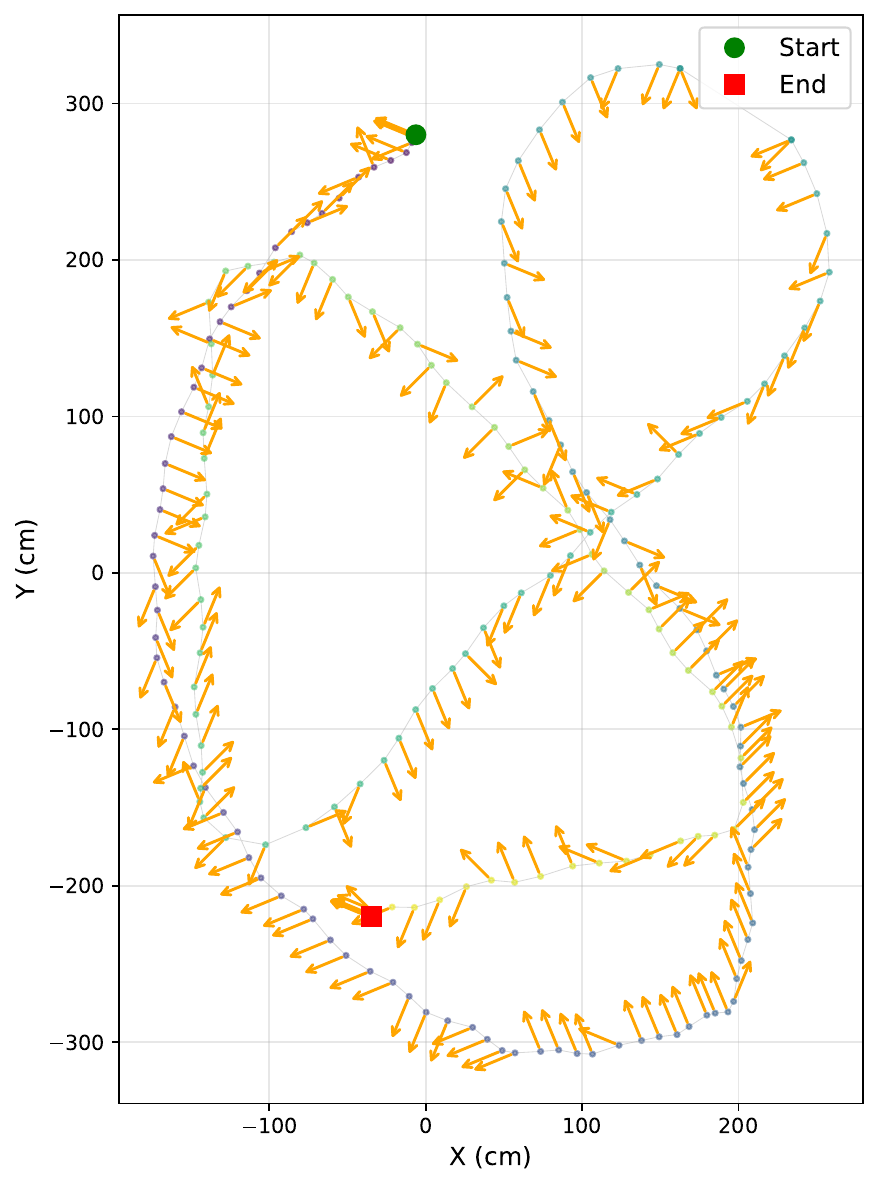}
        \caption{Rule-based}
        \label{fig:rule}
    \end{subfigure}

    \caption{Qualitative comparison of orientation estimation methods on one of the test sequences. Arrows indicate the predicted orientation direction at each position along the \gls{mocap} trajectory.}
    \label{fig:trajectory_comparison}
\end{figure*}

Examining the per-placement breakdown in Table~\ref{tab:combined_method_model_comparison}, the Arm placement yields the lowest raw \gls{mae} for both architectures (Mamba: $35.8^{\circ}$, Transformer: $40.8^{\circ}$), because arm-mounted tags experience more directional \gls{cir} variations during walking, while chest-mounted tags are the most challenging for the raw model. After applying Kalman filter, the pocket placement achieves the best overall results (Mamba: $13.0^{\circ}$, Transformer: $13.1^{\circ}$), attributable to the strong correlation between the walking direction captured by the position-derived heading and the orientation of a pocket-mounted tag.

\Rev{
\begin{table}[t]
    \centering
    \caption{Computational comparison of the two architectures.}
    \label{tab:complexity}
    \begin{tabular}{lcc}
        \toprule
        \textbf{Metric} & \textbf{Bi-Mamba} & \textbf{Transformer} \\
        \midrule
        Parameters & $409$\,K & $1.14$\,M \\
        Model size & $1.6$\,MB & $4.4$\,MB \\
        Sequence complexity & $\mathcal{O}(N_{\max})$ & $\mathcal{O}(N_{\max}^{2})$ \\
        \bottomrule
    \end{tabular}
\end{table}
}

Table~\ref{tab:complexity} compares the computational footprint of the two deep learning architectures. The bidirectional Mamba model contains $409$\,K trainable parameters and occupies $1.6$\,MB of storage, $2.8\times$ fewer parameters and $2.7\times$ smaller than the Transformer ($1.14$\,M parameters, $4.4$\,MB). In terms of theoretical complexity, Mamba's selective scan scales linearly with the sequence length $N_{\max}$ as $\mathcal{O}(N_{\max})$, whereas the Transformer's self-attention is $\mathcal{O}(N_{\max}^{2})$. This advantage is modest for the short anchor sequences used here ($N_{\max}=5$), but becomes significant in setups with denser anchor deployments.

Figure~\ref{fig:cdf} shows the cumulative distribution of absolute yaw errors across all Mamba test sequences. Location Kalman filter achieves the lowest median error ($p_{50} = 9.4^\circ$) and compresses the tail of the distribution: the 90th-percentile error drops from $114.4^\circ$ for the raw model to $45.0^\circ$, which is critical for safety-relevant applications. Orientation Kalman filter improves the 90th percentile compared to the raw model ($84.0^\circ$ vs.\ $114.4^\circ$) even though their medians are similar ($23.7^\circ$ vs.\ $22.4^\circ$), showing that temporal smoothing eliminates large errors rather than shifting the bulk of the distribution. The rule-based method exhibits the poorest tail behavior ($p_{90} = 158.6^\circ$), consistent with its coarse angular quantization.

Figure~\ref{fig:trajectory_comparison} provides a qualitative comparison on one representative test sequence. The ground-truth arrows (a) are spatially consistent and tightly aligned with the walking direction throughout the trajectory. Location Kalman filter (b) closely tracks the ground truth, with smooth and well-oriented arrows at nearly every position. Orientation Kalman filter (c) is noticeably smoother than the Mamba raw (d) but shows more drift in regions of rapid orientation change. The rule-based output (e) is coarse and erratic, with large angular jumps that render it unsuitable for fine-grained orientation tracking.

\section{Conclusion}
\label{sec:conclusion}

This work demonstrated that accurate human body orientation estimation can be achieved using only UWB channel impulse response measurements from a single wearable tag, eliminating the need for inertial sensors or multi‑tag configurations. Experimental results demonstrate that temporal modeling of CIR data using a Mamba neural network enables robust orientation inference (\gls{mae} of $38.6^\circ$), while Kalman‑based post‑processing significantly improves accuracy by exploiting motion continuity and position‑derived heading cues (\gls{mae} of $18.9^\circ$). Compared to a rule‑based baseline, the proposed approach reduces the mean absolute orientation error by more than 50\%. Importantly, the proposed pipeline operates reliably at low update rates and relies solely on UWB hardware that is already present in many indoor localization systems, making it well suited for practical deployments such as interactive museums, smart access control, and context‑aware indoor environments.


\bibliographystyle{unsrt}
\bibliography{References}

\end{document}